\input psfig
\documentstyle[twocolumn,aps,prl,floats]{revtex}
\begin{document}               
\title{The  $Q^2$ evolution of the generalized Gerasimov-Drell-Hearn integral for the neutron using a $^3$He target}
\author{
M.~Amarian,$^{5}$
L.~Auerbach,$^{20}$
T.~Averett,$^{6,23}$
J.~Berthot,$^{4}$
P.~Bertin,$^{4}$
W.~Bertozzi,$^{11}$
T.~Black,$^{11}$
E.~Brash,$^{16}$
D.~Brown,$^{10}$
E.~Burtin,$^{18}$
% ...
}
\address{
\baselineskip 7 pt
\vskip 0.1cm
{\rm The Jefferson Lab E94010 Collaboration} \break
\vskip 0.1 cm
{$^{1}$California Institute of Technology, Pasadena, California 91125} \break
{$^{2}$Chungnam National University, Taejon 305-764, Korea} \break
{$^{3}$Hampton University, Hampton, Virginia 23668} \break
{$^{4}$LPC IN2P3/CNRS, Universit\'e Blaise Pascal, F--63170 Aubi\`ere Cedex,
France} \break 
{$^{5}$Istituto Nazionale di Fiscica Nucleare, Sezione Sanit\`a, 00161 Roma, Italy}\break
{$^{6}$Thomas Jefferson National Accelerator Facility, Newport News, Virginia 23606} \break
{$^{7}$Kent State University, Kent, Ohio 44242}\break
{$^{8}$University of Kentucky, Lexington, Kentucky 40506}\break
{$^{9}$Kharkov Institute of Physics and Technology, Kharkov 310108, Ukraine} \break
{$^{10}$University of Maryland, College Park, Maryland 20742} \break
{$^{11}$Massachusetts Institute of Technology, Cambridge, Massachusetts 02139} \break
{$^{12}$University of Massachusetts Amherst, Amherst, Massachusetts 01003}\break
{$^{13}$University of New Hamphsire, Durham, New Hamphsire 03824} \break
{$^{14}$Old Dominion University,  Norfolk, Virginia 23529} \break
{$^{15}$Princeton University, Princeton, New Jersey 08544} \break
{$^{16}$University of Regina, Regina, SK S4S 0A2, Canada} \break
{$^{17}$Rutgers, The State University of New Jersey, Piscataway, New Jersey 08855 } \break
{$^{18}$CEA Saclay, DAPNIA/SPhN, F--91191 Gif sur Yvette, France} \break 
{$^{19}$Syracuse University, Syracuse, New York 13244} \break 
{$^{20}$Temple University, Philadelphia, Pennsylvania 19122} \break
{$^{21}$Tohoku University, Sendai 980, Japan} \break
{$^{22}$ University of Virginia, Charlottesville, Virginia 22904}\break
{$^{23}$The College of William and Mary, Williamsburg, Virginia 23187} \break
{$^{24}$Yerevan Physics Institute, Yerevan 375036, Armenia} \break
\break \break
} 
\maketitle
\begin{abstract}                
We present data on the inclusive scattering of polarized electrons from a polarized $^3$He target at energies from 0.862 to 5.06~GeV,
obtained at a scattering angle of 15.5$\,^{\circ}$.   Our data include measurements from the quasielastic peak, through the resonance
region, to the beginning of the deep inelastic regime, and were used to determine the virtual photon cross-section difference
$\sigma_{1/2}-\sigma_{3/2}$.   We extract the extended Gerasimov-Drell-Hearn integral for the neutron in the  range of 
4-momentum transfer squared $Q^2$ of  0.1--0.9~(GeV)$^2$.  
\end{abstract}

Sum rules involving the spin structure of the nucleon offer an important opportunity to study Quantum Chromodynamics (QCD).
At long distance scales or in the confinement regime, a sum rule of great interest is that due 
to Gerasimov, Drell and Hearn (GDH)\cite{ger,dre66}. The GDH sum rule relates an integral over the full excitation spectrum of
the spin-dependent total photoabsorption cross section to the nucleon's anomalous magnetic moment. It has not been investigated
experimentally until recently\cite{ahr01},  and further measurements are needed for a definitive
test.  At short distance scales or in the perturbative regime, two  other sum rules, one due to Bjorken\cite{bjo66} and the other
derived by Ellis and Jaffe\cite{ell74}, have provided us with significant information on nucleon spin structure through an extensive
experimental and theoretical investigation\cite{hug99}. These sum rules make predictions involving the first moments of the spin
structure functions measured in deep inelastic scattering (DIS).

The GDH sum rule pertains strictly to the real photon case for which the
four momentum transfer $Q^2$ =0. DIS data, in contrast, are taken at
relatively high values of $Q^2$. It is desirable to have both experimental
and theoretical bridges between these two very different regimes. This
was achieved by generalizing the ``GDH integral" to include the scattering of
virtual photons for which $Q^2>0$\cite{ans89}. The GDH integral is often written as
\begin{eqnarray}
I(Q^2) & = & \int_{\nu_0}^{\infty}{{d\nu}\over{\nu}}
\left[\sigma_{1/2}(\nu,Q^2)-\sigma_{3/2}(\nu,Q^2)\right]   \nonumber \\
 & = &  2\,\int_{\nu_0}^{\infty}{{d\nu}\over{\nu}}
\sigma^{\prime}_{TT} \ \ ,
\label{eq:gdhsum_def1}
\end{eqnarray}
where alternative generalizations are discussed in \cite{ji99} and \cite{dre00}.  Here $\sigma_{1/2\,(3/2)}(\nu,Q^2)$ is the total {\it virtual}
photoabsorption cross section  for the nucleon with a projection of
$1\over2$ ($3\over2$) for the total spin along the direction of photon momentum,  $\nu$ is the electron's energy loss,  $\nu_0$ is the
pion  production threshold, and 
$\sigma^{\prime}_{TT}$ is the  transverse-transverse interference cross section.  As $Q^2\rightarrow0$, $I(Q^2)$ is predicted by the
original GDH sum rule:
\begin{equation}
I(0) = -{{2\pi^2\alpha}\over{M^2}}\kappa^2,
\label{eq:gdh}
\end{equation}
where $\alpha$ is the fine structure constant, $M$ is the mass of the nucleon, and $\kappa$ is the nucleon's anomalous magnetic
moment.  As $Q^2\rightarrow\infty$, $I(Q^2) \rightarrow 16\pi^2\alpha\Gamma_1/Q^2$, where  $\Gamma_1=\int_0^1 g_1 dx$ is the
first moment of the nucleon's spin structure function $g_1$, and $x=Q^2/2M\nu$ is the Bjorken scaling variable.  
Both $\Gamma_1^p$  for the proton and $\Gamma_1^n$ for the neutron have been well studied  experimentally at high
$Q^2$\cite{hug99}.  At the low $Q^2$ limit, recent measurements have been made on $I(0)$ for the proton\cite{ahr01},
but the neutron has yet to be measured.  The two limits of the $Q^2$ evolution of  $I(Q^2)$ are thus constrained by a combination of
theory and experimental data.

It has recently been emphasized\cite{ji99} that the extended GDH integral can be related\cite{ji99,iof84,ber93} to the forward virtual
Compton scattering amplitudes thus establishing a true $Q^2$ dependent sum rule.  The original GDH sum
rule and the Bjorken sum rule can both be viewed as special cases of this ``extended GDH sum rule".  
We note that in the Bjorken sum rule the difference of $I^p(Q^2)$ for the proton and $I^n(Q^2)$ for the
neutron is considered.  The extended GDH sum rule can be tested at any value of $Q^2$ for which the Compton amplitudes can be
computed.   Near $Q^2=0$, where the amplitudes are best understood in terms of hadronic degrees of freedom, several calculations
have been performed using chiral perturbation theory ($\chi$PT)\cite{ji99,ji00,ber93,ber02}.    
The two most recent of these efforts, which take different approaches, include next-to-leading order corrections\cite{ji00,ber02}.  An
important issue is the highest values of $Q^2$ at which the calculations are valid, with estimates ranging as high as $0.3\,{\rm
GeV^2}$\cite{ber02}.
For large
$Q^2$, a region best described by partonic degrees of freedom, Operator Product Expansion (OPE)
techniques have been used to express the Compton amplitudes as a 
perturbative series in $\alpha_s$ and a (higher twist) power series in
$1/Q^2$.  The predictions of the Bjorken sum rule can thus be extended to finite $Q^2$, perhaps as low as 
$Q^2 \sim 0.5\,{\rm GeV}^2$\cite{ji00}.   For lower values of $Q^2$, that are still well above the range where $\chi$PT is applicable,
there is currently little theoretical guidance.  
This transition region, however,  is well suited to the use of lattice QCD\cite{isg00}.  Mapping
out $I(Q^2)$ experimentally, as we have done in this paper for the range of $0.1\,{\rm GeV}^2 \le Q^2 \le 0.9\,{\rm GeV}^2$, is an
important step to testing these ideas and building our understanding of the dynamics of nonperturbative QCD.

We measured the inclusive scattering of longitudinally polarized electrons from a polarized $^3$He target in Hall A of the Thomas
Jefferson National Accelerator Facility (JLab).  Data were collected at six incident beam energies: 5.06, 4.24,
3.38, 2.58, 1.72, and 0.86~GeV, all at a nominal scattering angle of 15.5$^{\circ}$.  The  measurements
covered values of the invariant mass $W$ from the quasielastic peak (not discussed in this paper), through the resonance region,  to 
the values indicated in Fig.~1.   Data were taken for both
longitudinal and transverse target polarization orientations.  Both spin asymmetries and absolute cross sections were measured.

A cw beam of  polarized electrons was produced by illuminating a strained GaAs photocathode with circularly polarized light. 
Beam currents were limited to  $10-15\,\mu\,$A to minimize depolarization of the target. The average polarization of the electron beam
was $0.70 \pm 0.03$, and was monitored using a double arm M\o ller polarimeter.  The polarization  was typically reversed at a rate of
1~Hz. 

The $^3$He target was polarized by spin-exchange with optically pumped rubidium (Rb)\cite{wal97}. Built specifically for this
and subsequent experiments in Hall A, the design of the target is similar to that built at SLAC for E142\cite{ant93},  but with
greater flexibility for polarization direction and increased capacity for handling higher average beam currents\cite{jen00,e94010}.  The
average in-beam polarization was $0.35 \pm 0.014$.  The
$^3$He gas was contained in sealed glass cells with densities corresponding to 10--12 atm~at 0$^{\circ}\,{\rm C}$.  
A small quantity (about 70~Torr) of nitrogen was also present to aid in the optical pumping process.  The portion of the cell in the electron
beam, the target chamber,  was a long cylinder approximately 40~cm in length.   The length was chosen so that the glass end windows
of the cell were at the edge of the acceptance of the spectrometers.  The portion of the cell in which optical pumping took place,
the pumping chamber, was irradiated with about 90~W of light from high-power diode-laser arrays with their wavelength centered at the
first transition line of Rb ($795\,{\rm nm}$).   The polarization of the
$^3$He was monitored by the NMR technique of adiabatic fast passage (AFP)\cite{abr61}.  The response of the NMR system was
calibrated by two methods. In one method the AFP signals of
$^3$He were compared with the AFP signals  from water which had a small polarization due to the usual Boltzmann distribution.  In the
other method AFP signals were observed following a determination of the $^3$He polarization using a shift of the Rb EPR
lines due to collisions with  polarized $^3$He atoms\cite{rom98}.  Independent verification of our target
polarimetry at the level of 4\% was provided by measuring the asymmetry in elastic scattering\cite{deu00,xu00}, which
depends on the product of the target and beam polarizations.

The  scattered electrons were detected using the two Hall A high resolution spectrometers.  Momenta were determined by
track analysis, and particle identification was accomplished using gas Cerenkov detectors and lead-glass shower counters.  Pion
rejection was better than $10^3$ in both spectrometer arms, which was more than sufficient since the $\pi/e$ ratio was 
never worse than 10.   

The quantities we measure experimentally are related to $\sigma^{\prime}_{TT}$ and the transverse-longitudinal interference term
$\sigma^{\prime}_{LT}$ in the Born approximation  according  to the relations
\begin{equation}
{{d^2\sigma^{\downarrow\Uparrow}}\over{d\Omega dE^{\prime}}}  - {{d^2\sigma^{\uparrow\Uparrow}}\over{d\Omega
dE^{\prime}}}= B\left( \sigma^{\prime}_{TT} + \eta\, \sigma^{\prime}_{LT} \right)
\label{eq:long_asym}
\end{equation}
and
\begin{equation}
{{d^2\sigma^{\downarrow\Rightarrow}}\over{d\Omega dE^{\prime}}}  - {{d^2\sigma^{\uparrow\Rightarrow}}\over{d\Omega
dE^{\prime}}}= B \sqrt{{2\epsilon}\over{1+\epsilon}}
\left( \sigma^{\prime}_{LT} - \zeta\, \sigma^{\prime}_{TT} \right)
\label{eq:tran_asym}
\end{equation}
where ${{d^2\sigma^{\downarrow\Uparrow(\uparrow\Uparrow)}}/{d\Omega dE^{\prime}}}$ is the cross section 
for the case in which the beam and target spin directions are antiparallel (parallel), and the 
left side of (\ref{eq:tran_asym}) represents the corresponding quantity for transverse target spin orientation. 
Also $B=2\,(\alpha/4\pi^2)(K/Q^2)(E^\prime/E)(2/(1-\epsilon))\left(1-{{E^{\prime} \epsilon}/{E}}\right)$,
where $E$ and $E^{\prime}$ are the initial and final electron energies, 
$\epsilon^{-1} = 1+2[1+Q^2/4M^2x^2]\tan^2(\theta/2)$, $\theta$ is the scattering angle in the laboratory frame, 
$\eta = \epsilon\sqrt{Q^2}/(E-E^\prime\epsilon)$, and
$\zeta=\eta(1+\epsilon)/2\epsilon$.  The factor $K$ represents the virtual photon flux and is convention dependent.  We use the
convention $K=\nu-Q^2/2M$, due to Hand\cite{han63}.

To extract from our raw $^3$He data $\sigma^{\prime}_{TT}$, which is defined within the Born approximation, we must first apply
``radiative corrections" to account for the emission of real and virtual photons.   These corrections were
performed using the procedure first described by Mo and Tsai for the case of unpolarized scattering\cite{mo69},  and extended to
include polarized effects using the program POLRAD\cite{aku94}.   For our experiment, we incorporated into POLRAD our actual data
for the quasielastic and resonance regions, as their effect on the radiative corrections are significant.  The results for
$\sigma^{\prime}_{TT}$ are shown in Fig.~1 as a function of the invariant mass $W$ for each of the six energies
measured, and represent the neutron to the extent that we have set $M$ equal to the neutron mass
in equations (3) and (4).  We note, though, that for  $\sigma^{\prime}_{TT}$,  no  corrections for nuclear effects were
applied.  
\begin{figure}[ht!]
\centerline{\ \psfig{figure=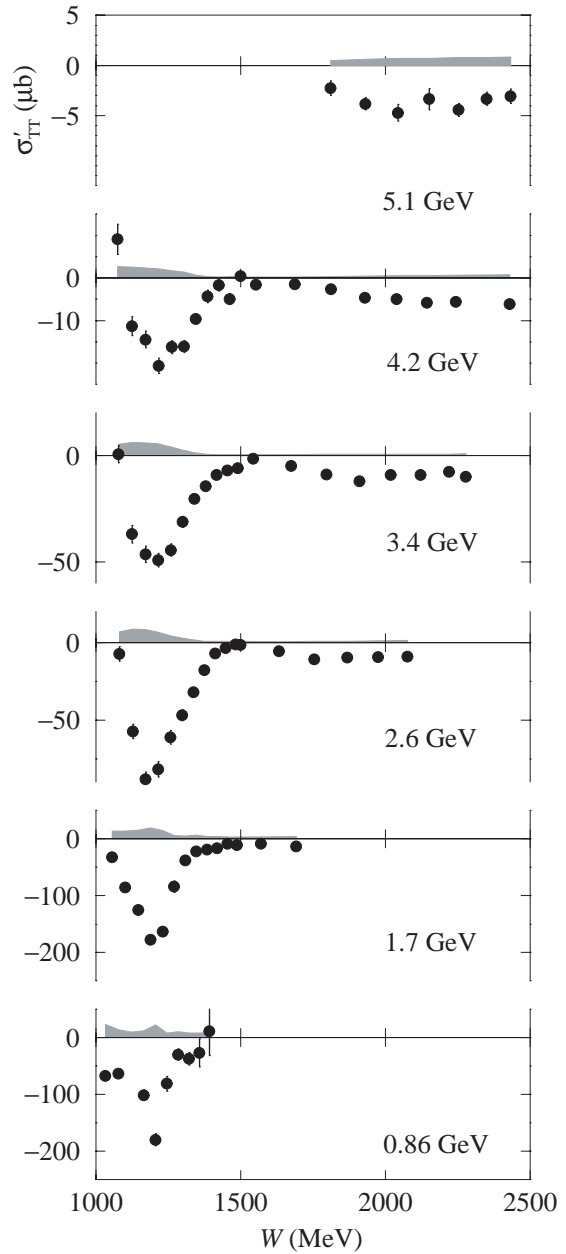,width=2.9in}}
\caption{$\sigma^{\prime}_{TT}$ is plotted as a function of invariant mass $W$ for each of the six incident energies studied.}
\label{fig:sigma_TT}
\end{figure}
The error bars are due to statistics only, with the grey bands indicating systematic errors.  The error due to radiative corrections, which
dominates the systematic errors on $\sigma^{\prime}_{TT}$,
was taken as 20\% of the applied correction for all but the lowest energy where 40\% was
assumed.  Other systematic errors included a relative
uncertainty in normalization of  5\%  from absolute cross section, 4\% from target polarization, and 4\% from beam polarization.  

To compute $I(Q^2)$, $\sigma^{\prime}_{TT}$ is needed at constant $Q^2$.  
We chose six equally spaced values of $Q^2$ in the range  $0.10\,{\rm GeV}^2 \le Q^2 \le 0.9\,{\rm GeV}^2$  and determined 
$\sigma^{\prime}_{TT}$ from our measured points by interpolation , or for a few points, extrapolation.
The results are plotted in Fig.~2 as a function of $\nu$.
\begin{figure}[ht!]
\centerline{\psfig{figure=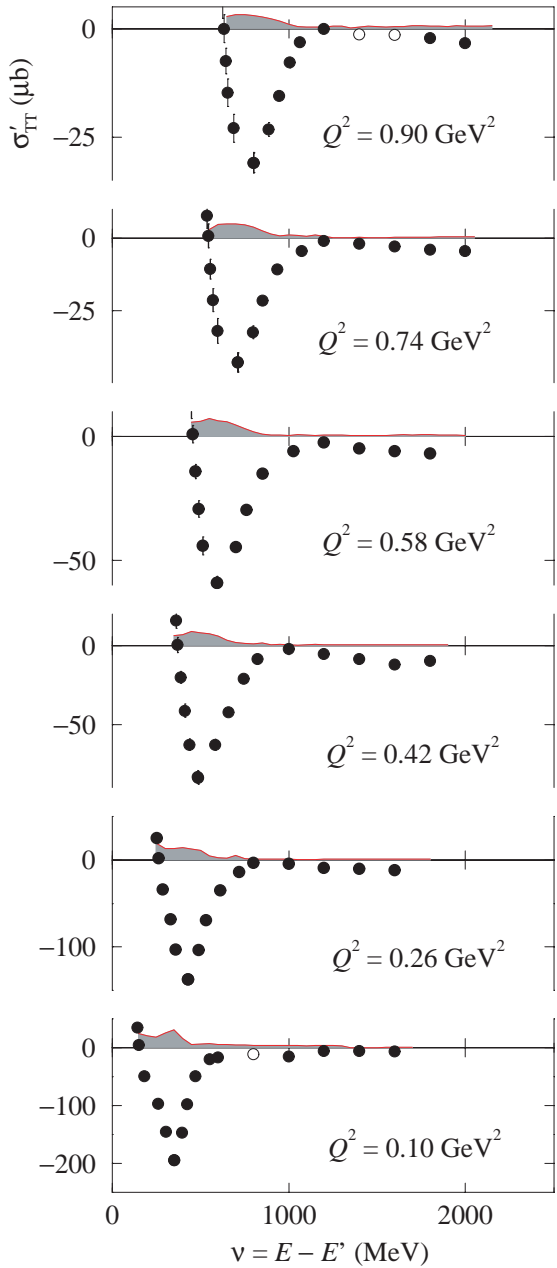,width=2.9in}}
\caption{$\sigma^{\prime}_{TT}$ is plotted as a function of energy loss $\nu$ for each of six values of constant $Q^2$.
The points shown with solid (open) circles were determined by interpolation (extrapolation).}
\label{fig:sigma_TT}
\end{figure}
The prominent peak in the cross section is the $\Delta_{1232}$
resonance,  which decreases in magnitude with increasing $Q^2$.  The error bars represent the uncertainty due to statistics, and
the grey bands indicate the uncertainty due to systematic errors, which in addition to those shown in Fig.~1, include a 
contribution from interpolation and extrapolation.

The extended GDH integral was computed for each value of $Q^2$ according to eq.~1  using limits of integration
extending from the nucleon pion threshold to a value of $\nu$ corresponding to $W=2.0\,{\rm GeV}$.  
The results are given in Table I.  Before plotting our results, we have applied a correction to account for the fact that our
neutron was embedded in a
$^3$He nucleus using a calculation due to
Ciofi degli Atti and Scopetta\cite{cio96}. This procedure introduces an additional 5\% systematic uncertaintly in our result.
Our results
for $I(Q^2)$ for the neutron, with the integration covering roughly the resonance region, are shown in Fig.~3 using open circles. 
The error bars, when visible, represent statistical uncertainties only, and the systematic effects are shown with the grey band.   
We have made an estimate of the unmeasured strength in
$I(Q^2)$ for the region
$4\,{\rm GeV^2} < W^2 < 1000\,{\rm GeV^2}$ using the parameterization of Thomas and Bianchi\cite{tho00} ($1000\,{\rm GeV^2}$
was the highest value considered in their paper). The solid squares have this estimate included, and an estimate of the theoretical
uncertainty has already been included in the systematic error shown.
\begin{figure}[ht!]
\centerline{\psfig{figure=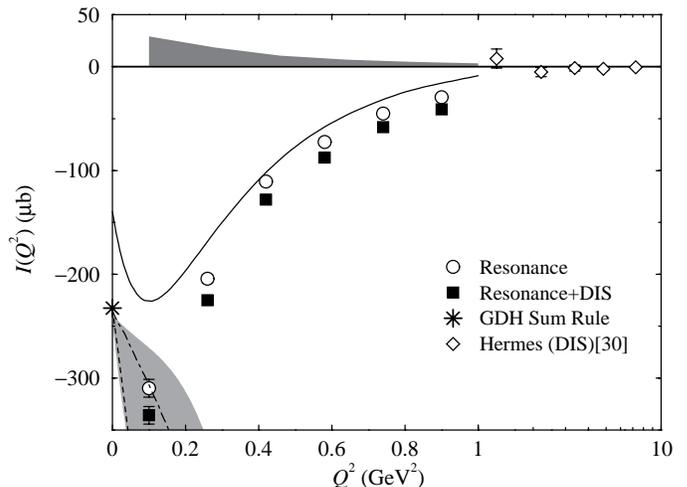,width=3.5in}}
\caption{Our measurements for $I(Q^2)$ vs. $Q^2$, both with and without an estimate of the DIS contribution.
Also shown with a dotted (dot-dashed) line are the $\chi$PT calculations of ref.~[12] (refs.~ [13] and [14]).  The calculation of ref.~[11],
based largely on the MAID model, is shown with a solid line.  We have included data from HERMES [30], and to avoid compressing our
horizontal scale, we have adopted a semi-log scale for $1\,{\rm GeV^2}<Q^2$.} 
\label{fig:coating}
\end{figure}

\begin{table}[ht!]
\begin{tabular}{|c|r|r|r|} 
$Q^2$ {(GeV$^2$)} & \multicolumn{1}{c|}{$I_{\rm GDH}$ {($\mu b$)}} &
\multicolumn{1}{c|}{Statistical {($\mu b$)}} & \multicolumn{1}{c|}{Systematic {($\mu b$)}} \\ \hline
0.10 & -223.76 & 6.18 & \multicolumn{1}{r|}{23.68} \\ \hline
0.26& -156.52 & 3.01 & \multicolumn{1}{r|}{13.58}\\ \hline
0.42& -88.73 & 2.10 &  \multicolumn{1}{r|}{7.46}\\ \hline
0.58& -59.56 & 1.51 &  \multicolumn{1}{r|}{4.56}\\ \hline
0.74& -37.96 & 1.43 &  \multicolumn{1}{r|}{3.14}\\ \hline
0.90& -23.63 & 1.00 &  \multicolumn{1}{r|}{2.09}\\ 
\end{tabular}
\caption{Measured values for $I(Q^2)$ prior to nuclear corrections together with statistical and systematic errors.}
\end{table}

Our data indicate a smooth variation of $I(Q^2)$  to increasingly negative values as $Q^2$ varies from $0.9\,{\rm GeV^2}$ 
toward zero.
Our data are more negative than the prediction of Drechsel, Kamalov,
and Tiator, whose calculation is mostly based on the phenomenological model MAID, and is shown on Fig.~3 as a solid 
line\cite{dre00}. Their prediction includes contributions to $I(Q^2)$ for $W \le 2\,{\rm GeV}$,  and should thus be compared with the
open circles.  At high
$Q^2$,  our data approach those of HERMES, which spans the range $1.28\,{\rm GeV}^2 < Q^2 < 7.25\,{\rm GeV}^2$\cite{ack98}, 
but includes only the DIS part of the GDH integral. It is desirable to compare our data with the GDH sum rule prediction
$I(0)=-232.8\,\mu{\rm b}$.  This prediction is indicated on Fig.~3, along with  extensions to $Q^2>0$ using two $\chi$PT
calculations, that described in \cite{ji00} (dotted line) and collectively in \cite{ber02} and \cite{mei02}(dot-dashed line).  
Taken together, \cite{ber02} and \cite{mei02} describe an extended calculation including resonance effects, and have associated
with them a large error (shown with a grey band) due to the uncertainty in certain resonance parameters.  The error can be seen to
encompass both our lowest $Q^2$ point and the calculation of \cite{ji00}.
We note, though, that at  $Q^2=0.3\,{\rm GeV^2}$, the prediction of \cite{ber02} and \cite{mei02} is much more negative than our data. 
We look forward to both further calculations as well as further measurements\cite{che97}. 

In conclusion we have made the first measurements of $\sigma^{\prime}_{TT}$ and the generalized GDH integral $I(Q^2)$ of the
neutron at low $Q^2$  ($0.1\,{\rm GeV}^2 \le Q^2 \le 0.9\,{\rm GeV}^2$).  
The data show a dramatic change in the value of the integral from what is observed at high $Q^2$ ($0.9\,{\rm GeV}^2<Q^2$).  While
not unexpected from phenomenological models, our data illustrate the sensitivity of $I(Q^2)$ to the transition from 
partonic to hadronic behavior.    Our data provide a precision data base for twist expansion analysis, provide a check
of $\chi$PT calculations, and  establish an important benchmark against which one can compare future calculations and 
measurements of low $Q^2$ spin structure.

We would like to thankfully acknowledge the untiring support of the JLab staff and accelerator division.  This work was supported by the
U.S. Department of Energy (DOE), the DOE-EPSCoR, the U.S. National Science Foundation, NSERC of Canada, the European INTAS
Foundation, the Italian INFN, and the French CEA, CNRS, and Conseil R\'egional d'Auvergne.   The Southeastern Universitites
Research Association (SURA) operates the Thomas Jefferson National Accelerator Facility for the DOE under contract
DE-AC05-84ER40150.

\end{document}